\documentclass[prl,twocolumn,showpacs,floatfix,amsmath,amssymb,superscriptaddress] {revtex4}
\usepackage{amssymb}
\usepackage{latexsym}

\usepackage{epsfig}

\usepackage{dcolumn}
\usepackage{amsmath}
\usepackage{amsfonts}
\usepackage{bm}

\newcommand{\be}{\begin{equation}}
\newcommand{\ee}{\end{equation}}
\newcommand{\bea}{\begin{eqnarray}}
\newcommand{\eea}{\end{eqnarray}}

\newcommand{\p}{\partial}
\newcommand{\s}{\sigma}

\newcommand{\la}{\langle}
\newcommand{\ra}{\rangle}
\newcommand{\rd}{\mbox{d}}
\newcommand{\ri}{\mbox{i}}
\newcommand{\re}{\mbox{e}}

\begin{document}
\title{Z$_N$ parafermion zero modes without Fractional Quantum Hall effect.}
\author{ A. M. Tsvelik}
\affiliation{ Department of Condensed Matter Physics and Materials Science, Brookhaven National Laboratory,
  Upton, NY 11973-5000, USA} \date{\today } 
\begin{abstract} 
I discuss a one-dimensional model of interacting fermions which collective excitations are Z$_N$-parafermions. The phase diagram of this model contains ground states with Charge Density Wave and superconducting quasi long range order. Boundaries between these phases contain zero energy parafermion modes.

 \end{abstract}


\maketitle
\section{Introduction.} 

Realization of quantum computation requires a possesion of the modern day philosophical stone - zero energy bound states with non-Abelian braiding statistics. The most discussed  among such bound states  are Majorana zero modes, but they are insufficient for universal quantum computation which requires more complicated, parafermion bound states (see, for example, \cite{kitaev},\cite{nayak},\cite{mong}). Permutation  of the anyons transforms one ground state into another one locally indistinguishable from the first \cite{bais},\cite{goldin}. In conformal field theories this property appears as  non-trivial braiding of the conformal blocks \cite{bzp} (see also \cite{knizhnik}). 
In quantum computation applications information is supposed to be stored nonlocally in  the zero energy modes and one has to learn how to manipulate them in order to process it.  These important problems have been discussed in the literature, but in the present paper I occupy myself only with the problem of existence of the anyon zero modes and nothing else. 

As platforms for non-Abelian anyons most  current theoretical schemes consider constructs based on  Fractional Quantum Hall effect systems (see \cite{mong} and references therein). The building blocks for such constructs are quantum wires placed in contact with a superconductor.  It is assumed that the bulk of the wires acquires a spectral gap via proximity-induced superconductivity and the edges contain counterpropagating Hall edge states. Such systems  require for their realization a strong magnetic field which does not square well with the superconducting pairing. Below I discuss a fermionic  model which  needs  neither magnetic field nor strong interactions.  I will demonstrate that the phase diagram of this model includes regions with Charge Density Wave and superconducting quasi long range order (LRO). The boundaries of their domains  contain zero energy modes of Z$_N$ parafermions \cite{fendley2}. This is the central result of the paper.

\section{The model.} 

In what follows I will consider  the model describing spin S=1/2 fermions with an additional orbital degeneracy. Its lattice version is 
\bea
&& H = \sum_r\Big\{ -t(\psi^+_{n,\alpha}(r+1)\psi_{n,\alpha}(r) + H.c.) +\nonumber\\
&& U[\psi^+_{n,\alpha}(r)\psi_{n,\alpha}(r)]^2 + \nonumber\\
&& g[\psi^+_{n,\alpha}(r)\epsilon_{\alpha\beta}\psi^+_{n,\beta}(r)][\psi_{m,\gamma}(r)\epsilon_{\gamma\delta}\psi_{m,\delta}(r)]\Big\}. \label{orig}
\eea
The interaction includes the standard Hubbard term $U$ and the pairing interaction $g$. The model was introduced in \cite{ho} in the context of cold atom physics. It is a particular case of the most general model describing fermionic atoms with half-integer spin $F$. The most general Hamiltonian with a point-like interaction contains $F+1/2$ parameters. I  consider the model far from  half filling $k_F \neq \pi/2$ (the Fermi wave vector) when the Umklapp processes are not important. 

 In the limit $U,g << t$ one can linearize the spectrum and replace the fermion operators by their slow components: 
\bea
\psi(r) = \re^{\ri k_Fr}R(r) +\re^{-\ri k_F r}L(r).
\eea
As a result we arrive to the field theory (relativistic) version of the Hamiltonian density: 
\bea
&& {\cal H} = -\ri R^+_{n,\alpha}\p_xR_{n,\alpha} + L^+_{n,\alpha}\p_x L_{n,\alpha}  - \nonumber\\
&& \Big( g_{cdw}O_{cdw}O^+_{cdw}  + g_{sc}O_{sc} O^+_{sc}\Big)\nonumber\\
&& + g_c(R^+_{n,\alpha}R_{n,\alpha})(L^+_{m,\beta}L_{m,\beta}),   \label{Model1}
\eea
where the Charge Density Wave and the superconducting order parameter operators are 
\bea
 O_{cdw} = (R^+_{n,\alpha}L_{n,\alpha}), ~~ O_{sc} = (R^+_{n,\alpha}\epsilon_{\alpha\beta}L^+_{n,\beta}).
\eea
and $g_{cdw} = -2U, ~~ g_{sc} = -4g, ~~ g_c = 2U$.  I have omitted terms containing fermion operators of the same chirality; they generate a renormalization of the Fermi velocity which I neglect. 

 The interaction  in (\ref{Model1}) contains both the superconducting pairing and the Charge Density Wave (CDW) order parameters which leads to a competition between these two types of order.  Although such feature is present for $N=1$ when Sp(2) = SU(2), its generalization for higher number of orbitals requires  Sp(2N) group. Such generalization allows for $1/N$-expansion which has been exploited in \cite{coleman1},\cite{coleman2}.  In \cite{phil,phil1} model (\ref{Model1}) has been studied by means of Renormalization Group theory and Conformal Field Theory. 

 At $g_{cdw}, g_{sc} >0$ the corresponding interaction terms in (\ref{Model1}) are relevant. The coupling $g_c$ is always marginal. Depending on what coupling is greater, $g_{cdw}$ or $g_{sc}$ the dominant fluctuations are either CDW or SC.   

\textit{Special case $g_{cdw} = g_{sc}$.---} A good starting point for our discussion is  the  special symmetric case $g_{cdw} = g_{sc} \equiv g_o$ when the model is integrable. The details of the solution are given in the Supplementary material. 
When the coplings are equal  the model acquires an additional symmetry; 
 the term with $g_o$ in (\ref{Model1}) can be written in the form of the current-current interaction of the Sp$_1$(2N) left- and right-moving currents 
\bea
J^a = R^+\hat t^a R, ~~  \bar J = L^+\hat t^a L, 
\eea
where $\hat t^a$ are sp(2N) generators. Sp(2N) group is a subgroup of SU(2N) which generators   change sign under time-reversal:
\be
\hat\epsilon(t^a)^T\hat\epsilon=- t^a.
\ee
It is convenient in this case to represent each label of SU(2N) by a pair of numbers $(n,\alpha)$ with $n=1,...N$ and $\alpha = \pm 1$ so that $\hat\epsilon$ is the antisymmetric tensor acting on the Greek indices. This brings us to the notations of the previous Section. The completness relation for the generators  is 
\bea
t^a_{\alpha\beta}t^a_{\gamma\eta} = \delta_{\alpha\gamma}\delta_{\beta\eta} - \epsilon_{\alpha\gamma}\epsilon_{\beta\eta}. \label{complete}
\eea
From (\ref{complete}) it follows that 
\bea
(R^+\hat t^a R) (L^+\hat t^a L) = -{\cal O}_{cdw}{\cal O}^+_{cdw} - {\cal O}_{sc}{\cal O}_{sc}^+ \label{compl}
\eea
 

The su$_N$(2) right moving currents are defined as 
\bea
j^3 = R^+_{n,\alpha}R_{n,\alpha}, ~~ j^+ = R^+_{n,\alpha}\epsilon_{\alpha\beta}R^+_{n,\beta}, ~~ j^- = R_{n,\alpha}\epsilon_{\alpha\beta}R_{n,\beta}. \label{SU(2)}
\eea
and to get the left moving ones one has to replace $R^+,R$ by $L^+,L$. 

 The key to the subsequent discussion  is the fact that  the symmetric point Hamiltonian (\ref{Model1}) can be written as a sum of commuting Hamiltonians. This can be proven using conformal embedding \cite{embed}. Conformal embedding defines "fractionalization rules" for breaking up free fermion Hamiltonians  into sums of commuting Hamiltonians of different critical models. The embedding we need in our case is 
\be
O_1(4nk) = Sp_k(2n)\oplus Sp_n(2k), 
\ee
which  $k=1$ version is 
\be
O_1(4N) = Sp_1(2N)\oplus SU_N(2). \label{embedding}
\ee
Applying (\ref{embedding}) to (\ref{Model1}) with $g_{cdw} = g_{sc} \equiv g_o$ and using (\ref{compl}) we arrive at the Hamiltonian density in the form of the WZNW models perturbed by a current-current interaction:
\bea
&& {\cal H} = {\cal H}_{Sp} + {\cal H}_{SU}, \\
&&{\cal H}_{Sp} = W[Sp_1(2N)] + \frac{g_o}{2}J_{ab}\bar J_{ba}, ~~ a, b = (n,\pm), \label{sp1}\\
&& {\cal H}_{SU} = W[SU_N(2)] + g_c j^3\bar{j}^3, \label{wzwsu}
\eea
where $W[G_k]$ is the Wess-Zumino-Novikov-Witten (WZNW) Hamiltonian density \cite{embed}.   For a Kac-Moody algebra G$_k$ it has a form
\bea
W[G_k] = \frac{2\pi}{k +c_v}\Big(:J^aJ^a: + :\bar J^a\bar J^a:\Big),
\eea
where $c_v$ is the quadratic Casimir in the adjoint representation (the Coexter number). Such models are integrable \cite{tsv87},\cite{ogievetskii},\cite{smirnov}. If the current-current interaction is relevant the excitations are non-Abelian anyons. For $G_k = sp(2N)$ their mass spectrum  is  
\bea
M_n = M \sin[\pi n/2(N+1)], ~~ n=1,...N.
\eea


\section{Attraction in the orbital channel. Competing orders. Emergent integrability. Z$_N$ parafermions. } 

My goal here is to consider small deviations from the integrability caused by inequality of the couplings $g_{cdw}, g_{sc}$. For this I have identify the most relevant operator associated with such deviation. My conclusions coincide with those of \cite{phil,phil1}, though I use a different arguments. 

 Let us consider the case $g_{cdw} = g_{sc}$ first. After Hubbard-Stratonovich transformation the interaction in the orbital channel becomes 
\bea
&& \frac{|\Delta_1|^2 + |\Delta_2|^2}{2g_o} + \label{HS}\\
&&\Big[\Big(\Delta_1 R^+_{n,\alpha}L_{n,\alpha} + H.c.\Big) + \Big(\Delta_2 R^+_{n,\alpha}\epsilon_{\alpha\beta}L^+_{n,\beta} + H.c\Big)\Big] \nonumber
\eea
At the saddle point we can approximate 
\bea
\Delta_a = |\Delta|z_a, ~~ \sum |z_a|^2 =1.
\eea
Then the fermion term becomes 
\bea
&& |\Delta|\Big(\bar\Psi_R \hat G\Psi_L + \bar\Psi_L\hat G^+\Psi_R\Big),~~\Psi_L = \left(
\begin{array}{c}
L_+\\
L^+_-\\
-L_-\\
L^+_+
\end{array}
\right), \nonumber\\
&& \bar\Psi = (\Psi^+)^T, ~~\hat G = \left(
\begin{array}{cc}
\hat g & 0\\
0 & \hat g
\end{array}
\right), ~~ \hat g = \left(
\begin{array}{cc}
z_1 & z_2\\
z_2^* & - z_1^*
\end{array}\right) \label{g}
\eea
Integration over the fermions yields the SU$_N$(2) WZNW model as it should be. The SU(2) matrix describes a combined order parameter which includes both CDW and SC. 

 Now let us consider what happens when the CDW and the SC pairing interactions have slightly different coupling constants $g_{cdw} - g_{sc} = \delta g_o$. In (\ref{HS}) this gives rise to the perturbation 
\be
V = \frac{\delta g_o}{g_o^2}|\Delta|^2 (|z_1|^2 - |z_2|^2) = \lambda \Phi_{zz}^{adj}, \label{relevant}
\ee
where $\Phi_{ab}^{adj}$ is the SU$_N$(2) primary field in the adjoint representation and $\lambda$ is its coupling constant. The operator field $\Phi_{ab}$ is $3\times 3$ matrix and the perturbation includes only one of its components. The perturbation is relevant, its scaling dimension is 
\be
d_{adj} =\frac{4}{N+2}.
\ee
This perturbation is integrable.  To see this I again use a conformal embedding
\be
SU_N(2) = U(1)\oplus Z_N, 
\ee
so that 
\bea
W[SU_N(2),g] + \lambda\Phi_{zz}^{adj} = \frac{N}{4\pi}(\p_{\mu}\phi)^2 + \Big[A[Z_N] + \lambda\Phi_{zz}^{adj}\Big], \label{ZN}
\eea
where $\phi$ is the U(1) phase and $A[{Z_N}]$ is the Lagrangian density of the critical parafermion theory. The field $\Phi_{zz}^{adj}$ is the thermal operator in the Z$_N$ model; the  latter theory perturbed by such operator remains integrable \cite{tsv88,fateev}.  Its spectrum consists of  massive excitations with a diagonal S-matrix 
\bea
[S(\theta)]_{a,b}^{\bar a,\bar b} = \frac{\sinh(\theta/2 + \ri\pi/N)}{\sinh(\theta/2 - \ri\pi/N)}\delta_a^{\bar a}\delta_b^{\bar b} \label{Sm}
\eea
and the mass spectrum given by 
\bea
&& m_n = m\frac{\sin(\pi n/N)}{\sin(\pi/N)}, ~~ n = 1,2,...N-1, \label{masses}\\
&& m \sim |(g_{cdw}- g_{sc})/(g_{cdw}+g_{sc})|^{1/(2-d_{adj})}M, \nonumber
\eea
 generated by the poles of (\ref{Sm}).

  The embedding suggests that one can express the components of the order parameter matrix  $g$ (\ref{g}) in terms of primary fields of (\ref{ZN}): 
\be
z_1 \sim \s \re^{\ri\phi}, ~~ z_2 \sim \mu\re^{\ri\theta},
\ee
where $\theta$ is the field dual to $\phi$ and $\s$ and $\mu$ are order and disorder parameters of the Z$_N$ parafermion model.  Depending on the sign of $\lambda$ it is either $\s$ or $\mu$ which acquires a vacuum expectation value which would correspond to either CDW (the  ordered phase of the parafermion model) or the SC phase.   The scaling dimensions of the corresponding bosonic exponents is $1/2N$.

\section{Parafermion zero modes.}

Now that we are in posession of parafermions, we can follow the standard idea of all works on zero modes (see, for example \cite{mong}).  Suppose that one can control the coupling constants $g_{cdw},g_{sc}$. Then one can create a heterostructure containing an array of zero energy parafermion modes similar to the one considered in \cite{mong}, but without a magnetic field. This heterostructure consists of  alternating domains with CDW or SC quasi long range order obtained by manipulation of the couplings. The zero modes reside on the domain boundaries. 

 To get an idea about possible observable effects of these modes we can look at the definition of the parafermions. Chiral parafermion creation and annihilation operators $\psi,\psi^+$ and $\bar\psi,\bar\psi^+$ are defined through the decomposition of the SU$_N$(2) fermionic currents (\ref{SU(2)})\cite{fatzam}:
\bea 
&& j^+ = \frac{\sqrt N}{2\pi}\re^{\ri\sqrt{8\pi}\varphi}\psi, ~~ j^- = \frac{\sqrt N}{2\pi}\re^{-\ri\sqrt{8\pi}\varphi}\psi^+\label{left}\\
&& j^z = \ri N/\sqrt{2\pi}\p_z\varphi\nonumber\\
&& \bar j^+ = \re^{-\ri\sqrt{8\pi}\bar\varphi}\bar\psi^+, ~~ \bar j^- = \re^{\ri\sqrt{8\pi}\bar\varphi}\bar\psi\label{right}\\
&& \bar j^z = -\ri N/\sqrt{2\pi}\p_{\bar z}\bar\varphi\nonumber,
\eea
where $\varphi, \bar\varphi$ are chiral components of the bosonic scalar field $\phi = \varphi + \bar\varphi$ governed by the Gaussian action (see (\ref{ZN})). From (\ref{left},\ref{right}) is clear that parafermion field are nonlocal, in the absence of interactions their correlation functions are not uniquely defined:
\bea
&& \la \psi(\tau,x)\psi^+(0,0)\ra \sim (\tau +\ri x)^{- 2(1-1/N)}, \nonumber\\
&&  \la \bar\psi(\tau,x)\bar\psi^+(0,0)\ra \sim (\tau -\ri x)^{- 2(1-1/N)}.
\eea
Presumably the zero modes retain this branch cut; then the local correlators of the parafermion fields will contain the time-independent piece corresponding to the zero mode:
\bea
&& \la \psi(\tau,x_1)\psi^+(0,x_2)\ra \sim \la \bar\psi(\tau,x_1)\bar\psi^+(0,x_2)\ra \\
&& \sim f(x_1)f^*(x_2) \exp[\ri\pi(1-1/N)\mbox{sign}\tau] + O(\re^{-m|\tau|}),\nonumber
\eea
where $f(x)$ is the zero mode wave function which decays exponentially with a distance from the domain wall.  As a consequence, the current-current correlation function at the domain wall will have  a nontrivial power law asymptotic:
\bea
\la j^+(\tau,x_1)j^-(0,x_2)\ra \sim \frac{f(x_1)f^*(x_2)}{|\tau|^{2/N}} + O(\re^{-m|\tau|}).
\eea
Observation of such behavior may serve as a signature of the zero mode. 


\section{Conclusions and Acknowledgements}

 The cold atom origin of model (\ref{orig}) gives us a hope that  its realization will not remain a  theorist's dream. Althought this may be a challenging task since for $N >2$ one requires fine tuning to arrive to  the necessary  form of the interaction, and only the case of $F=3/2$ fermionic atoms does not require this.  

I am grateful to A. Feiguin, H. Frahm and R. M. Konik for fruitful discussions. I am especially grateful to Ph. Lecheminant who attracted my attention to the relevant cold atom literature. The work was supported by the US DOE under contract number DE-AC02-98 CH 10886.

\section{Appendix. TBA equations for the Sp$_1$(2N) model}

 Thermodynamic Bethe ansatz (TBA) equations for model
 (13) can be extracted from \cite{ogievetskii}, \cite{og} which present a general solution for Bethe ansatz (BA) equations of integrable models with different Lie group symmetry. As a starting point for such derivation we can construct the coordinate Bethe ansatz for  the model with a linearized spectrum (3) using as a reference state the empty Fermi sea. Then to obtain the TBA equations in the relativistic limit (that is for model (13) one has to follow the well trodden path of filling the Fermi sea  and constructing the relativistic limit. The advantage of following this procedure instead of using the bootstrap \cite{zamzam} is that from the very beginning it deals with a nonrelativistic system and allows a detailed  study of the passage to the relativistic limit.  It turns out that the limiting procedure contains in this case an unexpected twist and therefore is rather instructive.  

Following \cite{ogievetskii} (see Eqs. (3.15) there ), \cite{og} I write down the discrete (BA) equations for model (3) with $g_{SC} = g_{CDW}$ when it has Sp(2N) symmetry: 
\bea
&& \prod_{b=1}^{M_{N-1}}e_4(x_a^{(N)} - x_b^{(N-1)}) =  \prod_{b=1}^{M_N}e_2(x_a^{(N)} - x_b^{(N)}),\nonumber\\
 &&\prod_{b=1}^{M_{p-1}}e_1(x_a^{(p)} - 
x_b^{(p-1)})\prod_{b=1}^{M_{p+1}}e_{1+\delta_{p,N-1}}(x_a^{(p)} - 
x_b^{(p+1)}) = \nonumber\\
&&\prod_{b=1}^{M_p}e_2(x_a^{(p)} - x_b^{(p)}),\quad p=2,\ldots,N-1,\nonumber\\ 
&& [e_1(x_a^{(1)}-1/\lambda)e_1(x_a^{(1)}+1/\lambda)]^{L/2}\prod_{b=1}^{M_2}e_1(x_a^{(1)} - 
x_b^{(2)}) = \nonumber\\
&&\prod_{b=1}^{M_1}e_2(x_a^{(1)} - x_b^{(1)}),\nonumber\\ 
&& E = \frac{1}{2i}\sum_a \ln[e_1(x_a^{(1)} -1/\lambda)/e_1(x_a^{(1)} +1/\lambda)]. \label{ON}
\eea
Here 
\[
e_n(x) = \frac{x-\ri n}{x + \ri n},
\]
$L$ is the number of particles in the Fermi sea, $\lambda$ is the bare coupling constant which determines the mass spectrum.

  In the thermodynamic limit imaginary parts of BA rapidities $x^{(p)}_a$ acquire fixed values and group into the so-called "strings". As a result  the discrete BA equations (\ref{ON}) may be replaced by equations for densities of the real  "string" solutions:
\bea
&& a_n\delta_{j1} = \tilde \rho_j^{(n)} + A_{nm}*C^{jk}*\rho_k^{(m)} -\delta_{j,N-1}A_{n,2m}*s*\s_m\nonumber\\
&& A_{2n,m}*s*\rho_{N-1}^{(m)} = \tilde\s_n + B_{nm}*\s_m
\eea
where $*$ stands for convolution 
\[
f*g(x) = \int \rd y f(x-y)g(y)
\]
and the Fourier transforms of the various kernels are given by 
\bea
&& a_n(\omega) = \cos(\omega/\lambda)\re^{-n|\omega|}, ~~ s(\omega) = [2\cosh(\omega)]^{-1}\nonumber\\
&& A_{nm}(\omega) = \coth(|\omega|)\Big[\re^{-|n-m||\omega|} - \re^{-(n+m)|\omega|}\Big],\nonumber\\
&& B_{nm}(\omega) = A_{nm}(2\omega), ~~ C_{jk}(\omega) = \delta_{jk} - s(\omega)[\delta_{j,k-1} + \delta_{j,k+1}].\nonumber
\eea

The corresponding TBA are 
\bea
&& T\ln(1+ \re^{\epsilon_n^{(j)}/T}) - TA_{nm}*C^{jk}*\ln(1+\re^{-\epsilon_m^{(k)}/T})  =\nonumber\\
&&  - \delta_{j,1}\tilde a_n - \delta_{j,N-1}TA_{n,2m}*s*\ln(1+\re^{-\kappa_m/T}),\nonumber\\
&&T\ln(1+ \re^{\kappa_n/T}) - TB_{nm}*\ln(1+\re^{-\kappa_m/T})  = \nonumber\\
&& - T*s*A_{2n,m}*\ln(1+ \re^{-\epsilon_m^{(N-1)}/T}), \label{TBAsp}\\
&& F/{\cal L}= - T\sum_{n=1}^{\infty}\tilde a_n*\ln(1+ \re^{-\epsilon_n^{(1)}/T})\nonumber
\eea
where ${\cal L}$ is the system size and 
\[
\tilde a_n(\omega) = \frac{\sin(\omega/\lambda)}{\omega}\re^{-n|\omega|/2}\epsilon_F, ~~ \epsilon_F = L/{\cal L}.
\]

Analizing the $T \rightarrow 0$ limit of (\ref{TBAsp}) one can see that the vacuum consists of real rapidities in the $N$-th equation  and real rapidities  and 2nd strings in all other equations. That is among all root densities only $\s_1, \rho_{1,2} \neq 0$. The Fourier transforms of the excitation energies are 
\bea
&&\epsilon_j^{(1)}(\omega) = \frac{\sinh(N-j)\omega}{\sinh(N\omega)}\frac{\sin(\omega/\lambda)}{\omega}, \nonumber\\
&& \epsilon_{j}^{(2)}(\omega) = \frac{\sinh j\omega}{2\sinh(N\omega)\cosh[(N+1)\omega]}\frac{\sin(\omega/\lambda)}{\omega}, \nonumber\\
&& \kappa_1(\omega) = \frac{1}{2\cosh[(N+1)\omega]}\frac{\sin(\omega/\lambda)}{\omega}.
\eea

Now comes the crucial point. In the relativistic limit  one should take the leading asymptotics of the excitation energies at $x=0$ which and here we have 
\bea
&& \epsilon^{(2)}_{j}(x)  =  M\sin\Big[\frac{\pi j}{2(N+1)}\Big]
\cosh[\pi x/2(N+1)], \nonumber\\
&& \kappa(x) \equiv \epsilon^{(2)}_N(x), \nonumber\\
&&\epsilon^{(1)}(x) = \Lambda\re^{-\pi/gN}\cosh(\pi x/N), \label{masses1}
\eea
where $M$ is the characteristic mass
\be
M = \Lambda \re^{-\pi/2\lambda(N+1)}.
\ee
From (\ref{masses1}) we see  that the spectral gap for $\epsilon^{(1)}$ is  $\sim \mu = \Lambda(M/\Lambda)^{2(N+1)/N}$. In the relativistic limit $M/\Lambda \rightarrow 0$ the parafermion dispersion it  vanishes, but for systems with finite coupling constants which we encounter in condensed matter physics it is not zero. Then $\mu$ is a finite energy scale and the physics of non-Abelian anyons emerges only above it. 
 In the first approximation  
\bea
&& \epsilon_2^{(j)}(\theta) = M_j\cosh\theta -h_j - \\
&& T\ln\Big\{\frac{\sin[\pi(j+1)/(N+2)]}{\sin[\pi/(N+2)]}\Big\} + O[\exp(- M_j/T)], \nonumber
\eea
where $h_j$ are magnetic fields coupled to the Kartan generators of Sp(2N). The fraction under the logarithm represents the quantum dimension of the j-the particle:
\be
d_j = \frac{\sin[\pi(j+1)/(N+2)]}{\sin[\pi/(N+2)]}.
\ee

 As a matter of fact, we can take another scaling limit, keeping $\mu$ constant. Then $M \rightarrow \infty$. In this limit $\epsilon_2$ must be dropped from the TBA equation for $\epsilon_1$ and as a result we get the TBA for the Z$_N$ model perturbed by the thermal operator:
\bea
&&T\ln(1+ \re^{-\epsilon_j^{(1)}(x)/T}) - {\cal A}_{jk}*T\ln(1+ \re^{\epsilon_k^{(1)}(x)/T})\label{tba1} \\
&& = m_j\cosh(\pi x/N) , ~~ m_j = m\frac{\sin(\pi j/N)}{\sin(\pi/N)}, ~~j=1,...N-1, \nonumber
\eea
where 
\[
{\cal A}_{jk} (\omega) = 2\coth\omega \frac{\sinh[(N - \mbox{max}(j,k))\omega]\sinh[\mbox{min}(j,k)\omega]}{\sinh(N\omega)}
\]
The same equations describe the situation of slightly unequal $g_{sc}, g_{cdw}$. This means that for the lattice model the position of the critical point $g_{sc}= g_{cdw}$ is shifted by the amount $\sim (M/\Lambda)$.

\end{document}